\documentclass[aps,
prl,
english,
reprint]{revtex4-1}

\usepackage{amsfonts,amsmath,amssymb}
\usepackage{graphicx}
\usepackage[utf8]{inputenc}
\usepackage{hyperref}
\usepackage{babel}

\newcommand\be{\begin{equation}}
\newcommand\ee{\end{equation}}
\newcommand\bea{\begin{eqnarray}}
\newcommand\eea{\end{eqnarray}}

\begin{document}

\def\rhoo{\rho_{_0}\!} 
\def\Zo{{Z_{_0}\!}}
\def\rhooo{\rho_{_{0,0}}\!} 

\begin{flushright}
\phantom{
{\tt arXiv:2106.$\_\_\_\_$}
}
\end{flushright}

{\flushleft\vskip-1.4cm\vbox{\includegraphics[width=1.15in]{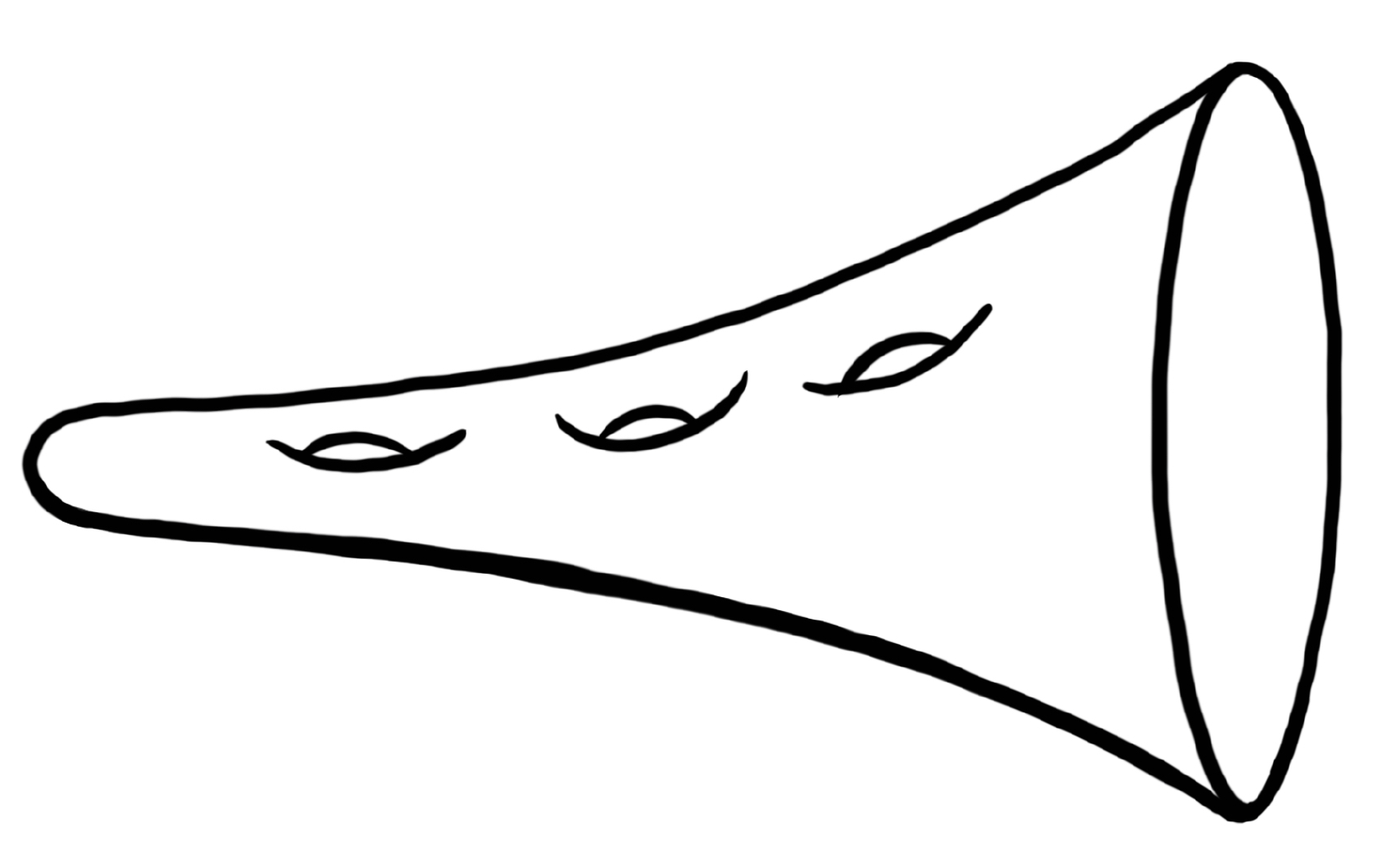}}}

\title{Quantum Gravity Microstates from Fredholm Determinants}
\author{Clifford V. Johnson}
\email{johnson1@usc.edu}
\affiliation{Department of Physics and Astronomy, University of
Southern California,
 Los Angeles, CA 90089-0484, U.S.A.}
 \date{June 17, 2021}


\begin{abstract}
A large class of two dimensional quantum gravity theories of Jackiw-Teitelboim form have  a description in terms of random matrix models. Such models, treated fully non-perturbatively,  can give an explicit and tractable description of the underlying ``microstate'' degrees of freedom. They play a prominent role in regimes where the smooth geometrical picture of the physics is inadequate. This is shown using a  natural tool for extracting the detailed microstate physics, a Fredholm determinant ${\rm det}(\mathbf{1}{-}\mathbf{ K})$. Its  associated kernel $K(E,E^\prime)$ can be  defined explicitly for a wide variety of JT gravity theories. To illustrate  the methods, the statistics of the first several energy levels of a non-perturbative definition of JT gravity are constructed explicitly using numerical methods, and the full quenched free energy~$F_Q(T)$ of the system is computed for the first time. These results are also of relevance to quantum properties of  black holes in higher dimensions.
\end{abstract}


\maketitle

%
%

{\it Introduction.---}Jackiw-Teitelboim (JT) gravity~\cite{Jackiw:1984je,*Teitelboim:1983ux} is a two dimensional model of gravity coupled to a scalar~$\phi$, with Euclidean action:
\begin{eqnarray}
\label{eq:JT-gravity-action}
I &=& - \frac12\int_{\cal M}\!\!\sqrt{g} \phi(R+2) -\int_{\partial \cal  M}\!\!\sqrt{h} \phi_b (K-1) \nonumber\\
&&\hskip 1.5cm -\frac{S_0}{2\pi}\left(\frac12\int_{\cal M} \!\!\sqrt{g} R +\int_{\partial {\cal M}}\sqrt{h}K\right)\ , 
\end{eqnarray}
where $R$ is the Ricci scalar and  in the  boundary terms,~$K$ is the trace of the extrinsic curvature for induced metric $h_{ij}$ and $\phi_b$ is the boundary value of $\phi$. The constant~$S_0$ multiplies the Einstein-Hilbert action, which yields the Euler characteristic $\chi({\cal M}){=}2{-}2g{-}b$ of the spacetime manifold ${\cal M}$ (with boundary $\partial {\cal M}$), with $g$ handles and $b$ boundaries. The partition function of the full quantum gravity theory $Z(\beta)$   at some inverse temperature $\beta{=}1/T$  is given by the path integral over ${\cal M}$ with a boundary of length $\beta$. It has a topological expansion $Z(\beta){=}\sum_{g=0}^\infty Z_g(\beta)$, where $Z_g(\beta)$ has a factor ${\rm e}^{-\chi({\cal M})S_0}$. 

JT gravity is of interest not just as a solvable toy model of  gravity, but also because it is a universal sector of the low $T$ near-horizon quantum dynamics of a wide class of higher dimensional black holes~\cite{Achucarro:1993fd,*Nayak:2018qej,*Kolekar:2018sba,*Ghosh:2019rcj}, including ones in four dimensional asymptotically flat spacetimes. $S_0$ is the extremal ($T{=}0$) Bekenstein-Hawking~\cite{Bekenstein:1973ur,*Hawking:1974sw} entropy and $\phi$ parameterizes the leading finite $T$ deviation of the geometry from extremality. Insights about this model therefore directly pertain to fundamental questions about quantum gravity and black holes in more ``realistic'' settings.

Solving the model~(\ref{eq:JT-gravity-action}) at leading (disc) order   by integrating out $\phi$ (locally enforcing $R{=}{-}2$ on ${\cal M}$)  results in all of the non-trivial dynamics being at the (length~$\beta$) boundary $\partial{\cal M}$. It has a Schwarzian action, and~\cite{Maldacena:2016hyu,*Maldacena:2016upp,*Jensen:2016pah,*Engelsoy:2016xyb}:
\be
\label{eq:density-on-disc}
\Zo(\beta) = \frac {{\rm e}^{S_0}{\rm e}^{\frac{\pi^2}{\beta}}}{4\sqrt{\pi}\beta^{\frac32}}\ , \,\,\,
\rhoo(E) = {\rm e}^{S_0}{\sinh(2\pi \sqrt{E})}/{4\pi^2}\ ,
\ee
 where the spectral density $\rhoo(E)$  comes from the Laplace transform: $\Zo(\beta){=} \int_0^\infty \rhoo(E) {\rm e}^{-\beta E}dE$. This class of models can be explicitly solved to any order to yield $Z_g(\beta)$ ({\it e.g.,} yielding corrections to $\rho_0(E)$), and correlations thereof.  This was shown in Refs.~\cite{Saad:2019lba,Stanford:2019vob}, along with a striking  equivalence (to be recalled below) to models of random $N{\times}N$ matrices at  large~$N$, where the topological expansion parameter ${\rm e}^{-S_0}{\sim}1/N$.

This Letter's results go well beyond the topological perturbative expansion  to uncover non-perturbative physics,   computing the details of  {\it individual underlying energy levels} of the spectrum. This  reveals the physics of the microscopic degrees of freedom of the quantum gravity theory, or, from the higher dimensional perspective, the underlying black hole microstates.  The entropy counts  microstates {\it via} $S_0{\sim}\log N$. Expanding in $1/N$ around large $N$ will not resolve their individual character. For this, a non-perturbative formulation is needed that allows $O({\rm e}^{-N})$ physics  to be extracted, a program begun in earnest (in this context) in Ref.~\cite{Johnson:2019eik}. The key new tool, introduced in this Letter, will be a  Fredholm determinant of an operator that  naturally arises from the underlying matrix model description.  It yields exquisite details of individual energy levels.

\begin{figure}[h]
\centering
\includegraphics[width=0.48\textwidth]{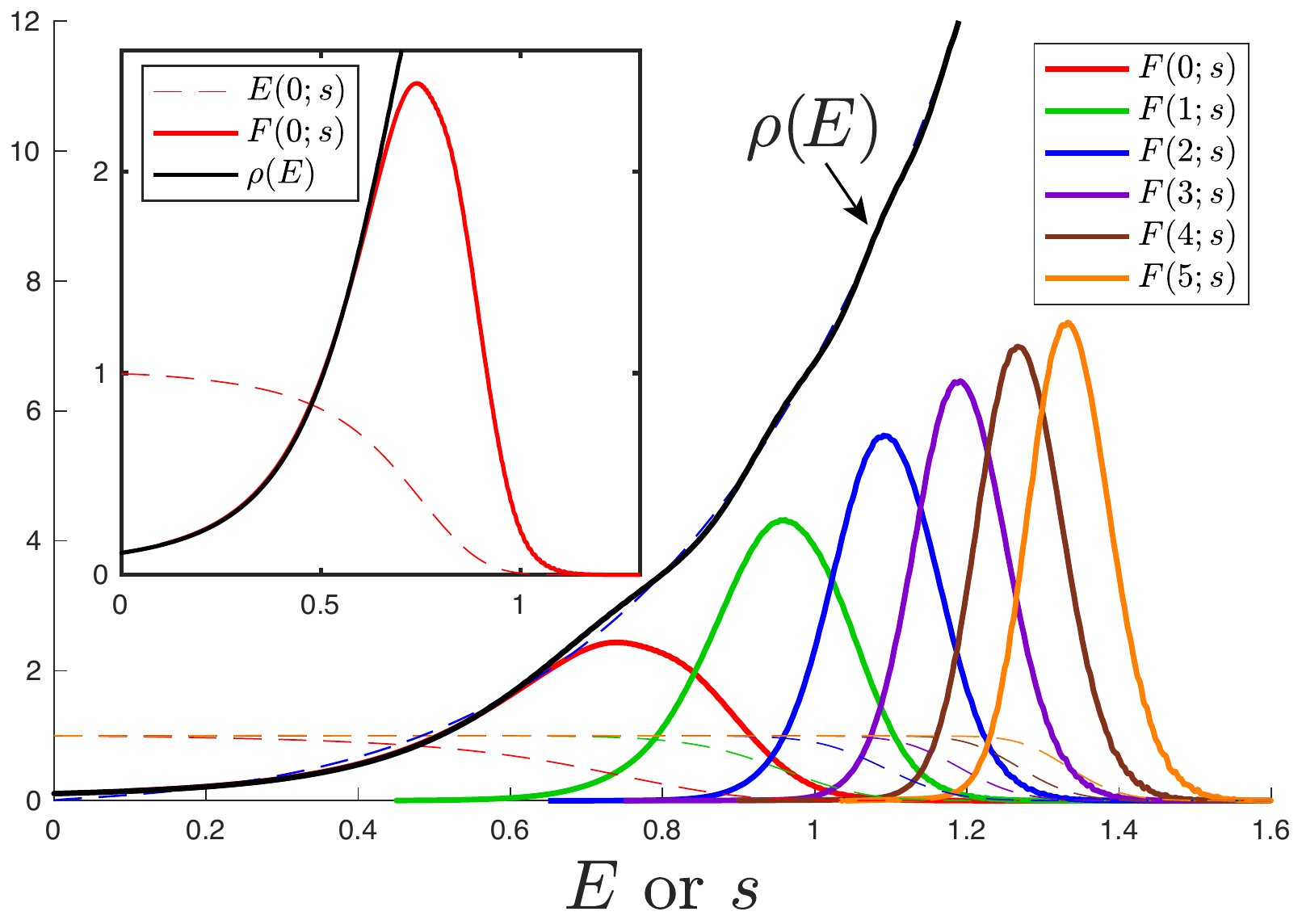}
\caption{\label{fig:JT-spectrum-from-fredholm} Spectral density $\rho(E)$ (solid black), $\rhoo(E)$ (blue dashed), and probability densities (also cumulative probabilities, dashed)  of the first 6  states of the  JT gravity microstate spectrum.
Inset: Close-up of $\rho(E)$ and  distributions for the ground state, with $\langle  E_0\rangle{\simeq}0.66$. Note that $\hbar{=}{\rm e}^{-S_0}{=}1$ here. }
\end{figure}

The results for the first six levels for JT gravity are shown in Fig.~\ref{fig:JT-spectrum-from-fredholm}. 
 %
For illustration, the plots (and all others herein) are  for ${\rm e}^{-S_0}{=}1$. For a fixed reference value of~$E$, smaller ${\rm e}^{-S_0}$ (larger extremal entropy $S_0$) results in an increase in the number of levels found to the left, which is nicely consistent with their    microstate interpretation.
A key feature is that the knowledge of the spectrum is {\it fundamentally statistical},  increasingly so at lower energies.~\footnote{This Letter will not dwell on the  (logically distinct) issue (discussed in Ref.~\cite{Maldacena:2004rf}, and now widely debated) of interpreting holographic correspondences when gravity seems to ensemble average over non-gravitational duals.} At higher $E$, the energy levels  become more sharply defined (their variance decreases), and also form a continuum. In this regime  the spacetime language (and the perturbation theory described above) is a good approximation. Low $E$ is  increasingly  non-perturbative, a spacetime interpretation falls short, and  the statistical description cannot be neglected. This suggests that geometry and statistics  are dual phenomena here. 

\begin{figure}[t]
\centering
\includegraphics[width=0.48\textwidth]{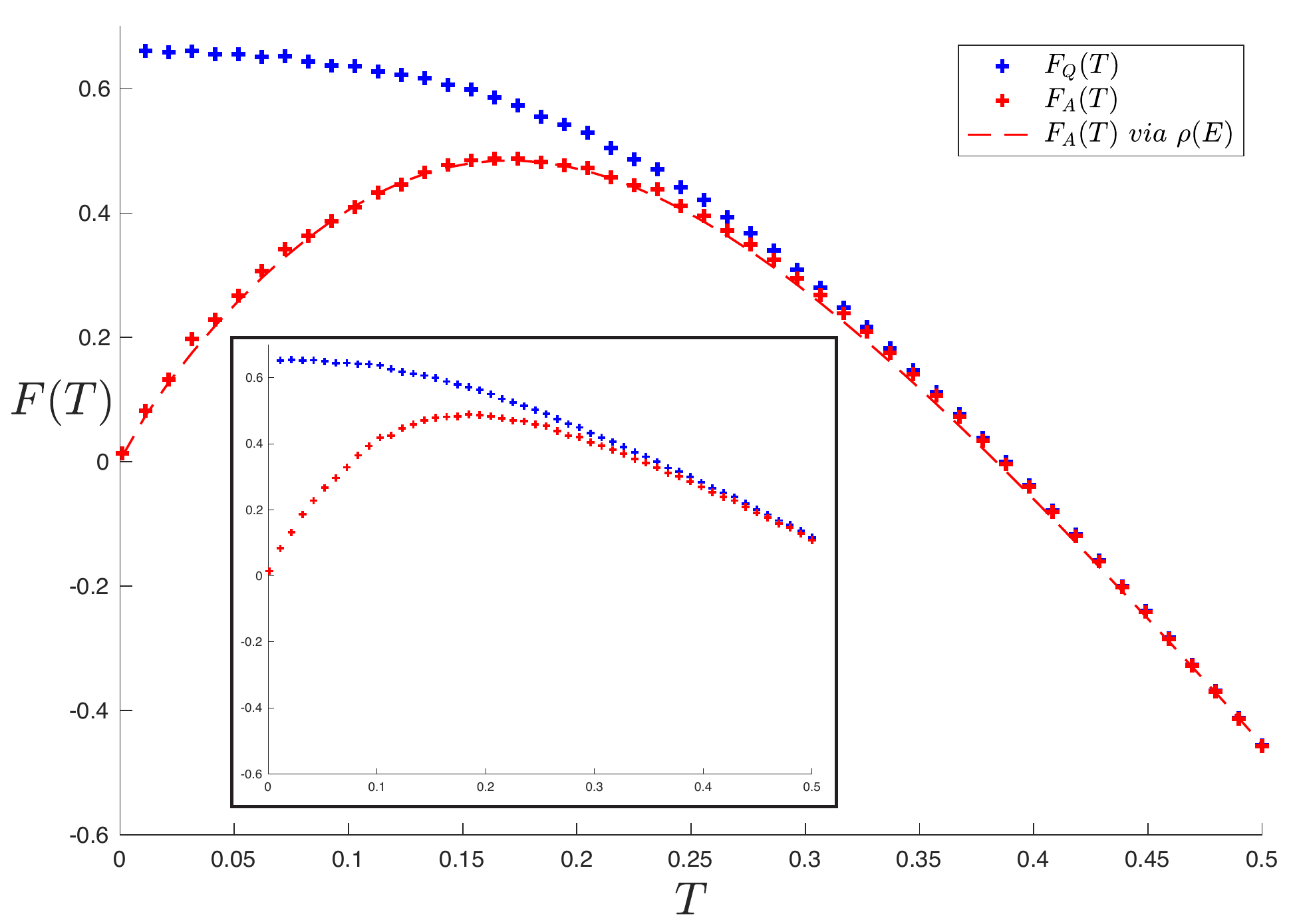}
\caption{\label{fig:JT-gravity-free-energy} The quenched and annealed free energies of JT gravity, computed using direct sampling (see text). As $T{\to}0$, $F_Q(T)$ lands at $\langle  E_0\rangle{\simeq}0.66$.  Here, $\hbar{=}{\rm e}^{-S_0}{=}1$.}
\end{figure}

Once the full spectrum and statistics are known,  any property of the model can be computed. A key example is the quenched free energy  $F_Q(T){=}{-}\beta^{-1}\langle\log Z(\beta)\rangle$, needed to compute thermodynamic quantities down to low $T$ (the more readily computable annealed free energy, $F_A(T){=}{-}\beta^{-1}\log \langle Z(\beta)\rangle$ gives a negative  entropy at low enough $T$). Its  computation has been discussed  frequently~\cite{Engelhardt:2020qpv,Johnson:2020mwi,Johnson:2021rsh}\cite{Okuyama:2020mhl,*Okuyama:2021pkf,*Janssen:2021mek}, but so far not completed.  Later in this Letter, the full spectrum will be used to explicitly compute  $F_Q(T)$ for the first time. It is displayed in Fig.~\ref{fig:JT-gravity-free-energy}.

{\it Perturbative Matrix Models.---}The full topological expansion of a large class of models of JT gravity  has been shown by Saad, Shenker and Stanford~\cite{Saad:2019lba} and Stanford and Witten~\cite{Stanford:2019vob} to  be captured~\footnote{This   includes variants with non-orientable ${\cal M}$, and also  supersymmetric extensions. Later work~\cite{Maxfield:2020ale,*Witten:2020wvy}  extended the connection to models with  more general potentials for~$\phi$.} by certain random matrix models in the ``double scaling limit'' of Refs.~\cite{Brezin:1990rb,*Douglas:1990ve,*Gross:1990vs}.   In the simplest matrix models, the probability distribution of the $N{\times}N$ matrix $M$ is $p(M){=}{\rm e}^{-{\rm Tr} V(M)}$ with  Gaussian $V(M){=}\frac{1}{2}M^2$ being the most famous prototype~\cite{10.2307/2331939,*10.2307/1970079}. 
A more general polynomial $V(M){=}\sum_p g_p M^p$ is of interest for studying gravity. (See {\it e.g.} Ref.~\cite{Ginsparg:1993is} for a  review.) Treating the matrix partition function ${\widetilde Z} {=}\int p(M)dM $ as  a toy field theory and working at large~$N$, the Feynman diagrammatic expansion can be viewed (following 't Hooft, and Brezin {\it et.~al.}~\cite{'tHooft:1973jz,*Brezin:1978sv}) as tessellations  of 2D Euclidean spacetimes, with each order in the $1/N$ expansion corresponding to the topology upon which the diagrams  can be drawn. The double scaling limit is a combination of sending $N{\to}\infty$ while also tuning the couplings~$g_p$ to critical values such that  surfaces large compared to the scale of the tessellation dominate. This yields universal continuum physics.  In this sense the matrix model can be thought of as an alternative method for doing the 2D gravity path integral. 
%
However, as already mentioned, fundamental quantum gravity questions will require non-perturbative physics---indeed going beyond geometry. A path integral definition that is intrinsically based on integrating  over smooth manifolds ${\cal M}$ cannot get access to this physics. Matrix models can unlock  this non-perturbative sector, since the geometrical spacetime regime  is only an ``emergent'' subsector of the theory.

 In the large $N$ limit the distribution of eigenvalues $\lambda_i$ becomes  dense and can be described with a smooth coordinate $\lambda$. The non-zero eigenvalues fill a finite segment of the real $\lambda$ line, and a spectral density $\rho(\lambda)$ describes their distribution. (The focus here will be on $V(M)$ that produce a single connected component.) The Laplace transform of $\rho(\lambda)$ is (roughly) the expectation value of a ``loop operator''  $\langle {\rm Tr} [{\rm e}^{\ell M}]\rangle$, which makes a loop of length~$\ell$ in the tessellations. The double-scaling limit ($N{\to}\infty$ with $g_p{\to}g^c_p$)  focuses on the scaling region in the neighbourhood of an endpoint~\cite{Dalley:1991zs,*Bowick:1991ky} (at, say, $\lambda{=}\lambda_0$), blowing up that region and sending the other end to infinity, {\it i.e.,} $\lambda{=}\lambda_0{+}E\delta^p$, as parameter $\delta{\to}0$, for some positive $p$. Finally,  the spectral density is some $\rho(E)$ supported on the real line starting at $E{=}0$ 
 stretching to infinity. The details of $\rho(E)$ depend upon the gravity theory in question.

In general, the dictionary between the choice of gravity theory and double-scaled potential is not straightforward, but in genus perturbation  theory, everything at higher genus is determined by the leading (disc) order results by ``topological recursion'' properties studied by Mirzakhani, and by Eynard and Orantin, equivalent to certain ``loop equations'' for correlation functions~\cite{Mirzakhani:2006fta,*Eynard:2007fi}. Crucially,  Ref.~\cite{Saad:2019lba} showed that the JT gravity partition function $Z(\beta)$ is equivalent to  a loop expectation  (of length~$\beta$).  At leading/disc order, the spectral density is set as Eq.~(\ref{eq:density-on-disc}), and the topological expansion parameter match: ${\rm e}^{-S_0}{\sim}1/N$ (after double-scaling).  They then checked that order by order in topology, the correlation functions of the matrix model that follow from the leading result equal the explicit results obtained by directly doing the gravitational path integral on manifolds~${\cal M}$. It is highly non-trivial and remarkable that it works.
Nevertheless, this is still all perturbative. The next step is to study non-perturbative contributions to~$\rho(E)$, beyond the reach of the recursion approach. 

{\it Non-Perturbative Matrix Models.---}The direct non-perturbative completion of Ref.~\cite{Saad:2019lba}'s  Hermitian matrix model definition of JT gravity has instabilities. One way to see this~\cite{Saad:2019lba,Johnson:2019eik} is to note that perturbatively  it is  a combination~\cite{Okuyama:2019xbv}  of  the family of $(2k{-}1,2)$ minimal models (coupled to gravity), obtained~\cite{Brezin:1990rb,*Douglas:1990ve,*Gross:1990vs} as Hermitian matrix models: It was known long ago that the $k$ even models are non-perturbatively unstable  due to eigenvalue tunneling to infinite negative values. However Ref.~\cite{Johnson:2019eik} supplied a  non-perturbative completion of the JT gravity matrix model that preserves the perturbative expansion to all orders.  It is built by double-scaling {\it complex} matrices~$M$ for which the potential is built from~$MM^\dagger$~\cite{Morris:1991cq,*Dalley:1992qg,*Morris:1992zr}. It can  be thought of as a model of an Hermitian matrix with a   manifestly positive eigenvalue spectrum. The model is unafflicted by non-perturbative instabilities.~\footnote{Other non-perturbative completions are proposed in {\it e.g.,} Refs.~\cite{Saad:2019lba,Gao:2021uro}.}

A useful way of seeing the difference between the two types of model is through their orthogonal polynomial formulation. The Hermitian matrix model integral can be written entirely in terms of a system of $N$ polynomials $P_i(\lambda)$,  orthogonal with respect to the measure $d\lambda {\rm e}^{-V(\lambda)}$. 
The polynomials themselves are related according to $\lambda P_i(\lambda){=}P_{i+1}(\lambda) {+} R_iP_{i-1}(\lambda)$  and the $R_i$ satisfy a recursion relation  determined by $V(\lambda)$. Knowing the coefficients $R_i$ turns out to be equivalent to solving the partition function integral or  any insertion into the integral  of traces of powers of $M$ 
{\it i.e.,} of $\lambda$.    In the large $N$ limit, the index $i/N$ becomes a continuous coordinate $X$ that runs from 0 to~1, the endpoints of the density. The orthogonal polynomials become functions of $X$: $P_i(\lambda){\to} P(X,\lambda)$, as do the recursion coefficients: $R_i{\to}R(X)$. Recall that the double scaled limit focuses on an endpoint, say $X{=}0$. In the limit, scaled versions of all the key quantities survive to define the continuum physics. For example, $X$ has a scaled piece~$x$, which runs over the whole real line, {\it via}: $X{=}0{-}x\delta^{p_x}$ where $p_x$ is some positive power and $\delta{\to}0$ in the limit. The topological expansion parameter $1/N$ also picks up a scaling piece, denoted  $\hbar$, {\it via}: $1/N {=} \hbar\delta^{p_n}$. For the recursion coefficients $R(X) {=} R(0){+}u(x)\delta^{p_u}$, with $p_u$ some other positive power. In the limit (the  powers $p_x,p_u$ {\it etc.,} are determined by requiring finite physics),  the recursion relation for $R_i$ becomes a {\it differential equation} for $u(x)$. It contains both perturbative and non-perturbative information about the model. The orthogonal polynomials $P_i(\lambda)$ become   $\psi(x,E)$,   wavefunctions of an Hamiltonian that is  the scaled part of~$\lambda$ (acting as an operator inside the integral): $\lambda{=}\lambda_0{+}{\cal H}\delta^p$, where: 
\be
\label{eq:hamiltonian}
 {\cal H}=-\hbar^2\frac{\partial^2}{\partial x^2} + u(x)\ .
\ee
Solving the double scaled matrix model boils down to determining $u(x)$. This in turn defines a quantum mechanics and   wavefunctions $\psi(E,x)$ and energies $E$ {\it via}: 
\be
\label{eq:eigenstates}
{\cal H}\psi(E,x)=E\psi(E,x)\ .
\ee
In fact, the loop operator in this language is~\cite{Banks:1990df}:
\be
\label{eq:full-loop-expression}
Z(\beta)  = \int_{-\infty}^0 \!\!dx \,\langle x| {\rm e}^{-\beta{\cal H}}|x\rangle\ ,
\ee a  trace over the exponentiated Hamiltonian. Putting $\int d\psi |\psi\rangle\langle\psi|{=}1$ into Eq.~(\ref{eq:full-loop-expression}) and using Eq.~(\ref{eq:eigenstates}) with $\psi{\equiv}\langle\psi|x\rangle$
gives $Z(\beta){=}\int dE \rho(E) {\rm e}^{-\beta E}$ with:
 \be
 \label{eq:full-density}
 \rho(E)=\int_{-\infty}^0|\psi(E,x)|^2dx\ ,
 \ee
{\it i.e.,} knowing the complete $\psi(E,x)$ can be used to construct the full spectral density to all orders and beyond.

The difference between the two types of model can now be explained, using the simplest (Gaussian) case as an example. In the Hermitian matrix model, the unscaled $P_i(\lambda)$ are  Hermite functions, 
and in the scaling limit at an endpoint  become~\cite{Moore:1990cn,*FORRESTER1993709} Airy functions $\psi(E,x){=}\hbar^{-\frac23}{\rm Ai}(-(E+x)\hbar^{-\frac23})$, which satisfy Eq.~(\ref{eq:eigenstates}) for potential $u(x){=}{-}x$. Their energies  range over the whole real $E$ line. 
Using them in Eq.~(\ref{eq:full-density}) gives the full non-perturbative density: 
$\rho(E)_{\rm Ai}{=}\hbar^{-\frac23}\left( {\rm Ai}^\prime(\zeta)^2{-}\zeta {\rm Ai}(\zeta)^2\right)$, with $\zeta{\equiv}{-}\hbar^{-\frac23}E$,  where ${f}^\prime{\equiv}\partial f/ \partial x$. Its support includes  $E{<}0$.
Large $E$ gives  the classical (disc order) result $\rhoo{=}E^\frac12/(\pi\hbar)$ supported only on  $E{\ge}0$. 
In contrast, for the complex matrix models, similar steps~\cite{Dalley:1992qg} give   a non-linear equation for   $u(x)$:
\be
\label{eq:cmm-string-equation}
u{\cal R}^2-\frac{\hbar^2}{2} {\cal R} {\cal R}^{\prime\prime}+\frac{\hbar^2}{4}({\cal R}^\prime)^2 = 0\ ,
\ee
where ${\cal R} {=} u(x){+}x$. 
The solution of interest 
has $u(x){=}{-}x$ for $x{\to}{-}\infty$ and $u(x){=}0$ for $x{\to}{+}\infty$. So for large energies the behaviour of the wavefunctions (in the $x{<}0$ regime, which corresponds to perturbation theory) is again like the Airy form of the Hermitian case, and so the physics   is perturbatively identical. However at low energies the wavefunctions become less Airy-like (in fact they join on to a Bessel-like behaviour at positive $x$), and moreover the energies in the problem are bounded below by zero. The resulting fully non-perturbative spectral density for this problem resembles  that of the Airy problem, but  naturally truncates~\cite{Johnson:2019eik}  at $E{=}0$. The non-perturbative states at $E{<}0$ are not present, while the perturbative physics remains the same. There is an analogous  model for any~$k$  with this kind of  behaviour, and Ref.~\cite{Johnson:2019eik} showed how to build a non-perturbative completion of JT gravity using them. It is a matter of finding the correct  equation for $u(x)$. It is constructed as follows: The general double scaled matrix model yields  the form~(\ref{eq:cmm-string-equation}) but   with  ${\cal R}{=}\sum_{k=1}^\infty t_k R_k[u] +x$, where $R_k[u]$  are the ``Gel'fand-Dikii''\cite{Gelfand:1976A} polynomials in $u(x)$ and its derivatives, beginning as $R_k[u]{=}u(x)^k+\cdots$.  (Their details are not needed here.)
The  $t_k$ are couplings. The specific values $t_k{=}\pi^{2k-2}/k!(k{-}1)!$ yield an equation that  to leading order gives the JT spectral density~(\ref{eq:density-on-disc}), with $\hbar{=}{\rm e}^{-S_0}$. 

Solving the equation for $u(x)$  order by order (expanding about large $x{<}0$ perturbation theory) yields the identical physics that Ref.~\cite{Saad:2019lba}  constructed. Meanwhile, the full solution for $u(x)$ (constructed numerically to good accuracy in Ref.~\cite{Johnson:2020exp}) determines~${\cal H}$. The spectral problem can be solved  numerically to yield  wavefunctions $\psi(E,x)$, and  the integral~(\ref{eq:full-density}) yields the non-perturbative~$\rho(E)$. It is plotted in Fig.~\ref{fig:JT-spectrum-from-fredholm} as the solid black curve. 
 At large~$E$ it asymptotes to the classical Schwarzian result~(\ref{eq:density-on-disc}) (dashed line), but at small~$E$ there are undulations that are invisible in perturbation theory.~\footnote{This method yielded explicit non-perturbative $\rho(E)$ for a large class of supersymmetric models in Refs.~\cite{Johnson:2020heh,Johnson:2020exp,Johnson:2020mwi}, as well as deformed JT gravity models and an additional supersymmetric model in Refs.~\cite{Johnson:2020lns,*Johnson:2021owr}.}   They are the precursors of  the underlying microphysics.

{\it Detailed Microphysics.---}A new type of tool will now be introduced that will take the construction much further than  before.   Fredholm determinants and their associated kernels are a powerful tool that have been used in the statistical physics literature for computing random matrix model properties such as correlation functions~\cite{Meh2004,*ForresterBook}. The determinantal structure  arises because of the van der Monde determinant $\prod_{i\neq j}(\lambda_i-\lambda_j)^2$ that comes as the Jacobian for going from $M$ variables  to eigenvalues $\lambda$. The orthogonal polynomial basis inherits all this structure, and can be used as  the building blocks, in their guise as the  $\psi(E,x)$.  For example the core object  is the  ``kernel'': 
\bea
\label{eq:kernel}
K(E,E^\prime)&=&\int_{-\infty}^0\psi(E,x)\psi(E^\prime,x) dx\ .
\eea
In fact, it  has already played a role, as its diagonal is the non-perturbative density, Eq.~(\ref{eq:full-density}). Clearly there is a lot more information to be extracted  from its off-diagonal terms. Various combinations of the kernel, Laplace transformed, can be used to write multi-point correlation functions~\cite{Banks:1990df} for  $Z(\beta)$, but the raw form will be a much more direct tool for studying the spectrum.

It is  the context of the Fredholm determinant construction that  gives $K$ its kernel moniker. In solving  problems of the form~\cite{10.1007/BF02421317}:
$
f(E) {-}\!\int_a^b K(E,E^\prime)f(E^\prime) dE^\prime{=} g(E)
$,
on some interval $(a,b)$ (or union of intervals) in the $E$--plane, 
 the properties of ${\rm det}(\mathbf{I}-\mathbf{K})$ are important.  $\mathbf{K}$ is the integral operator with kernel $K(E,E^\prime)$. 
The random matrix model literature is filled with various kinds of kernel of interest, (Airy,  Bessel, sine, Laguerre, {\it etc.,}). This Letter's studies of JT gravity now add some new  well-defined kernels to the menagerie. They may be of wider mathematical interest since ({\it e.g.,} for the $k{=}1$ model) the wavefunctions are a hybrid of Airy and Bessel functions, and $k$th generalizations thereof,  giving a portmanteau of Wigner and Wishart behaviour, scaled. 


Returning to the physics, the Fredholm determinant can be used to focus on one energy/eigenvalue at a time  as follows. For the $1$st energy level (labelled henceforth as the 0th, for the ground state, $E_0$, of JT gravity system), the probability of finding  no eigenvalues on the interval $(0,s)$ (chosen since the lowest possible energy  is zero) is $E(0;s){=} {\rm det}(\mathbf{I}-\mathbf{K}|_{(0,s)})$. Crucially (for later) this is a cumulative probability density function (CDF).  The probability density  function (PDF) for finding  an energy is  $F(0;s){=}{-}dE(0;s)/ds.$ For orientation, in the case of the Airy model the interval would be $(-\infty, s)$, and using the Airy kernel  yields the famous Tracy-Widom distribution~\cite{Tracy:1992rf} of the smallest eigenvalue. Here, the JT gravity system will reveal a new kind of smallest eigenvalue distribution. 

More generally,  for the $n$th energy level, the probability (CDF) for  the interval $(0,s)$  is written as:
\be
\label{eq:n-th-level-cdf}
E(n;s) =\sum_{j=0}^{n}\frac{(-1)^j}{j!}\frac{d^j}{dz^j} {\rm det}(\mathbf{I}-z\mathbf{K}|_{(0,s)})\Biggl.\Biggr|_{z=1}\ .
\ee Correspondingly the PDF is: $F(n;s){=}{-}d E(n;s)/ds$.
 

The  major challenge  now is to compute the determinant of the infinite dimensional operator in Eq.~(\ref{eq:n-th-level-cdf}). This requires much more care, and is prone to severe numerical difficulties even though the problem is effectively discrete ($\sim${~}700 energies were used). Useful at this point is the impressive work of Bornemann~\cite{Bornemann_2009}  that shows how to use quadrature on a relatively small number of points in the energy interval $(0,s)$ to compute Fredholm determinants: Using Clenshaw-Curtis quadrature to break up the interval into~$m$ points $e_i$ and compute weights $w_i$ ($i=1\cdots m$), an integral $\int_0^s f(E)dE $ on the interval would be computed as $\sum_i^m w_i f(e_i)$. So similarly the determinant becomes:
 \be
 {\rm det}(\mathbf{I}-z\mathbf{K}|_{(0,s)}) \to {\rm det}(\delta_{ij}-zw_i^\frac12 K(e_i,e_j)w_j^\frac12)\ .
 \ee
If the $\psi(E,x)$ are known accurately (such as for Airy and Bessel kernels)  the method  gives impressive results with modest values of $m$ such as 8, or 16. In the case in hand, the $\psi(E,x)$ were found as approximate numerical solutions to an eigenstate problem for which the potential $u(x)$ was itself a solution to a difficult (15th order~\cite{Johnson:2020exp}) non-linear equation, so some challenging numerical difficulties due to errors can be expected. However, they can be surmounted well enough to get very good results for the definition of JT gravity discussed here~\footnote{The {\tt Chebfun} package in {\tt MatLab}  helped produce more refined  $\psi(E,x)$ than were previously used in Refs.~\cite{Johnson:2019eik,Johnson:2020exp,Johnson:2020mwi}, giving a numerically cleaner $K(E,E^\prime)$.}. 

The result for the zeroth level (the ground state) is the focus of  the inset of  Fig.~\ref{fig:JT-spectrum-from-fredholm}. A small amount of  smoothing has been applied to remove numerical noise.
%
 To the left, at zero, the CDF $E(0;s)$  shows that there is a non-zero (but small) chance of finding the ground state there, falling steadily to zero to the far right with the increasing unlikelihood of finding the ground state at very high energies. Its derivative, the PDF $F(0;s)$ is also shown in the inset. It peaks at around 0.75.  The mean of the distribution gives the average ground state of the ensemble $\langle E_0\rangle{\simeq}0.66$. (For illustration purposes all computations were done using $\hbar{=}1$). Computing the results for higher levels is straightforward, although numerical inaccuracies in finding the first level get successively amplified with each level. The first six levels are shown in the main part of Fig.~\ref{fig:JT-spectrum-from-fredholm}. What has been uncovered here with the Fredholm determinant technique are the explicit probability peaks for individual energies/microstates that, when added together, produce the previously found non-perturbative undulations in the spectral density (black line). In principle any quantity can now be computed using this information, as will be demonstrated next.

{\it Quenched Free Energy.---}It is important to compute the quantity $F_Q(T){=}{-} \beta^{-1}\langle\log Z(\beta)\rangle$ for JT gravity, but it is difficult. Ref.~\cite{Engelhardt:2020qpv} pointed out that connected diagrams with multiple boundaries (``replica wormholes'', as they were implementing a gravitational  replica trick)  should play a  crucial role. Ref.~\cite{Johnson:2020mwi} observed that in addition a non-perturbative formulation was needed, such as a matrix model. There have been various useful partial results from toy models~\cite{Okuyama:2020mhl,Okuyama:2021pkf,Janssen:2021mek,Johnson:2021rsh}, at low $T$. 
Ref.~\cite{Johnson:2021rsh} noted that a better tool was needed, in order to incorporate the properties of individual peaks. The Fredholm determinant has performed handily in that regard, and so  computation  of $F_Q(T)$ is now rather straightforward. 

 A numerical approach is natural (since all the explicit data about the levels are numerical),  simply directly sampling ensembles. In fact, this  was done recently in Ref.~\cite{Johnson:2021rsh} for  the comparatively simple cases of the Airy model and a variety of Bessel models. There, all that was needed was Gaussian random matrices (readily numerically generated) plus scaling  to an endpoint. A matrix model for JT gravity has much more exotic probability distributions however, so direct sampling seems doomed. However, the individual probability distributions for each level  can now be generated by reverse-engineering the above results for the $E(n;s)$, which (recall) are  CDFs. A key result from the theory of statistics is that {\it any} probability distribution function can be generated from a uniform probability distribution by mapping from the associated CDF.  So, ensembles can  be  generated as follows:  For a particular sample, generate  the $n$th energy level $E_n$ with the appropriate probability (using uniform PDFs on a computer and converting using the  CDF $E(n;s)$ to sample the correct PDF $F(n;s)$).  Then compute $\log(Z(\beta)){=}\log(\sum_n {\rm e}^{-\beta E_n})$. This was done for an ensemble of just 5000 samples, and {\it then}  averaged. For contrast, the annealed quantity $F_A(T){=}{-}\beta^{-1}\log \langle Z(\beta)\rangle$ can be computed too (averaging the partition function over the ensemble and taking the log at the end). 
 
 Using just the first six levels yields the content of the inset of Fig.~\ref{fig:JT-gravity-free-energy}.  
As higher levels are added, the details of the curves settle swiftly. Only successively higher $T$ details are affected by adding successively higher levels, the process asymptoting to the classical result at large $T$. In the main plot of Fig.~\ref{fig:JT-gravity-free-energy}, a result with 150 levels is given. It is constructed by approximating the $E(n;s)$ for  levels above about $n{=}10$.  In this regime  the approximation is already very much under control. The point is that even by 6  levels (see Fig.~\ref{fig:JT-spectrum-from-fredholm}), the peaks have  narrowed and overlap significantly, as the system returns to the continuum (classical) regime. The $n{>}10$ peaks are well approximated by narrow Gaussians, with their location, mean,  and standard deviation  determined by the classical density curve~(\ref{eq:density-on-disc}). So CDFs  (error functions for Gaussians) were used  for high  levels. As a test of  this, the red dashed line is the annealed free energy computed using the $Z(\beta)$ obtained by simply Laplace transforming $\rho(E)$ (integrated to the appropriate cutoff energy corresponding to the 150th level). The agreement  with the result computed by direct ensemble averaging (red crosses) is remarkably good. This shows that the result for $F_Q(T)$ is accurate.
Overall, rather nicely,   $F_Q(T)$ is monotonically decreasing  ({\it i.e.,} the entropy $S(T)$ is manifestly positive), with zero slope at $T{=}0$, corresponding to entropy $S_0$ at extremality. Additionally, this result confirms an earlier suggestion~\cite{Johnson:2020mwi} that  there is no sign of the replica symmetry breaking transition conjectured in Ref.~\cite{Engelhardt:2020qpv}.


{\it Final Remarks.---}The studies of this Letter (with  several JT supergravity examples to appear)
show that  matrix models of JT gravity  are completely tractable models of a key quantum gravity phenomenon: A cross-over from smooth, classical spacetime geometry (large $E$  (or~$T$) in Fig.~\ref{fig:JT-spectrum-from-fredholm} (or~\ref{fig:JT-gravity-free-energy})) to a regime where spacetime is not enough, and a description in terms of discrete microstates takes over (small $E$  (or $T$)). The microstates are always there, of course, but the regime where their individuality is inevitable is when the variance or randomness becomes pronounced, while inter-spacing is large. This is a regime beyond smooth spacetimes, even with many handles and wormholes. It is no longer geometrical, but the matrix model computes the physics just as readily. The key tool used here was the Fredholm determinant, whose kernel is built from sewing together two copies of the wavefunctions $\psi(E,x)$. Interestingly,  the $\psi(E,x)$ have an interpretation~\cite{Maldacena:2004sn} as a type of D-brane probe, in the language of minimal string theory. In a sense then, the Fredholm tool is a D-brane probe that detects an increased  spreading out or variance (or itself spreads out)  as it moves to lower energies. This is reminiscent of aspects of the D-brane probes involved in the ``enhan\c con mechanism''~\cite{Johnson:1999qt}. This might be relevant when considering what lessons these phenomena might teach about higher dimensional quantum gravity. Finally, as mentioned in the introduction, the microstates uncovered in detail here  also  model  the microstates of the higher dimensional  black holes whose near-horizon low~$T$ dynamics is controlled by a JT gravity. It will be interesting to see what other features of JT gravity and black hole physics (in various dimensions) might be accessible using matrix model technology.  

\smallskip
 
 \begin{acknowledgments}
CVJ  thanks  the  US Department of Energy for support under grant  \protect{DE-SC} 0011687, and, especially during the pandemic,  Amelia for her support and patience.    
\end{acknowledgments}

\bibliographystyle{apsrev4-1}
\bibliography{NP_super_JT_gravity,LE_super_JT_gravity}

\end{document}